\documentclass[aps,prb,preprint,showpacs,amsmath,amssymb]{revtex4-1}

\usepackage{amssymb}
\usepackage{amsmath}
\usepackage{bm}% bold math
\usepackage{color}

\usepackage[pdftex]{graphicx} % is important

\begin{document}

% Use the \preprint command to place your local institutional report number 
% on the title page in preprint mode.
% Multiple \preprint commands are allowed.
%\preprint{}

\title{Velocity autocorrelation in liquid para-hydrogen by quantum simulations for first-principle computations of the neutron cross sections} %Title of paper

% repeat the \author .. \affiliation  etc. as needed
% \email, \thanks, \homepage, \altaffiliation all apply to the current author.
% Explanatory text should go in the []'s, 
% actual e-mail address or url should go in the {}'s for \email and \homepage.
% Please use the appropriate macro for the type of information

% \affiliation command applies to all authors since the last \affiliation command. 
% The \affiliation command should follow the other information.

\author{E. Guarini$^{1}$, M. Neumann$^{2}$, U. Bafile$^{3}$, M. Celli$^{3}$, D. Colognesi$^{3}$, E. Farhi$^{4}$, Y. Calzavara$^{4}$}
%\email[]{guarini@fi.infn.it}
%\homepage[]{Your web page}
%\thanks{}
%\altaffiliation{}
\affiliation{$^{1}$ Dipartimento di Fisica e Astronomia, Universit\`a degli Studi di
Firenze, via G. Sansone 1, I-50019 Sesto Fiorentino, Italy\\ $^{2}$ Fakult\"{a}t f\"{u}r Physik der Universit\"{a}t Wien, Strudlhofgasse 4, A-1090 Wien, Austria\\ $^{3}$ Consiglio Nazionale delle Ricerche, Istituto dei Sistemi Complessi, via Madonna del Piano 10, I-50019 Sesto Fiorentino, Italy\\$^{4}$ Institut Laue-Langevin, 71 avenue des Martyrs, CS 20156, F-38042 Grenoble Cedex 9, France}

% Collaboration name, if desired (requires use of superscriptaddress option in
%\documentclass).
% \noaffiliation is required (may also be used with the \author command).
%\collaboration{}
%\noaffiliation

\date{\today}

\begin{abstract}

Accurate knowledge of the single-molecule (self) translational dynamics of liquid para-H$_2$  is an essential requirement for the calculation of the neutron scattering properties of this important quantum liquid. We show that, by using Centroid Molecular Dynamics (CMD) quantum simulations of the velocity autocorrelation function, calculations of the total neutron cross section (TCS) remarkably agree with experimental data at the thermal and epithermal incident neutron energies where para-H$_2$ dynamics is actually dominated by the self contributions. This result shows that a proper account of the quantum nature of the fluid, as provided by CMD, is a necessary and very effective condition to obtain the correct absolute-scale cross section values directly from first-principle computations of the double differential cross section, and without the need of introducing any empirically adjusted quantity. At subthermal incident energies, appropriate modeling of the para-H$_2$ intermolecular (distinct) dynamics also becomes crucial, but quantum simulations are not yet able to cope with it. Existing simple models which account for the distinct part provide an appropriate correction of self-only calculations and bring the computed results in reasonable accord with TCS experimental data available until very recently. However, if just published cross section measurements in the cold range are considered, the agreement turns out to be by far superior and very satisfactory. The possible origin of slight residual differences will be commented and suggest further computational and experimental efforts. Nonetheless, the ability to reproduce the total cross section in the wide range between 1 and 900 meV represents an encouraging and important validation step of the CMD method and of the present simple algorithm.

\end{abstract}

% insert suggested PACS numbers in braces on next line

\maketitle %\maketitle must follow title, authors, abstract and \pacs

% Body of paper goes here. Use proper sectioning commands. 
% References should be done using the \cite, \ref, and \label commands
%\section{}
%\label{}
%\subsection{}
%\subsubsection{}

\section{Introduction}

Hydrogen isotopes H$_2$ and D$_2$ in their liquid states have always attracted much interest for their importance as fundamental and simplest molecular systems. The ``hydrogens'' display a quantum behavior still challenging theoretical and simulation-based descriptions of their translational dynamics, as far as both single-molecule and collective properties are concerned. The study of these liquids is, therefore, evidently important {\it per se}; nevertheless, its relevance is even further enhanced by two facts. One is the direct access to their microscopic dynamics enabled by important and widespread spectroscopic techniques. Neutron scattering, in particular, is by far the most important means for studying the hydrogens and hydrogen-containing materials, and it is well known that the study of the cross section for scattering of neutrons from a sample of liquid hydrogen has been, by itself, a long-standing theme in the physics of quantum fluids (see e.g. [\onlinecite{guariniDDCS}] and Refs. therein). The second fact is that liquid hydrogens are among the most used cryogenic fluids and, in the specific application to neutron techniques, are the most important low-temperature neutron moderators used to realize cold neutron sources. 

In this field of application, an improved description of the neutron double differential cross section (DDCS) and total cross section (TCS) of liquid H$_2$ and D$_2$, and of other moderating materials, has become an indispensable requirement for appropriate development and upgrade of neutron facilities. For instance, it has been recently shown \cite{Farhi} that calculations of the safety rod insertion impact on the criticality of water-moderated reactors provide much better results if based on new accurate (experimental and/or simulated) DDCS determinations, rather than on existing cross section libraries and scattering kernels employed in the available nuclear data processing codes treating water and other neutron moderators.\cite{MacFarlane94,MattesKeinert,NJOY2012} Similarly, the capability of predicting, with a high reliability, the dynamic response to neutrons of the hydrogen liquids is of crucial importance for the design of advanced and high brilliance cold neutron sources, as those aimed to exploit directional moderator geometries. \cite{mezei, batkov} 

Direct experimental investigations of the scattering law of these systems are of course necessary to validate any model calculation, but obviously cannot cope with the present need to know the DDCS in a large number of conditions of exchanged wave vector $Q$ and energy $E$ for any relevant incident energy $E_0$. Therefore the development of computable accurate models for the DDCS is a major objective in this field of research and application, especially if focused on first-principle methods aimed to obtain direct agreement with experimental data without forcing it through the introduction of variable parameters or {\it ad hoc} values for some physical quantities. Unfortunately, available data, appropriately normalized to an absolute scale, mainly refer to integrated quantities like the TCS measurements of Refs.\ [\onlinecite{seiffert,celli1999,grammer2015}], while past DDCS determinations for hydrogen and deuterium are often given in arbitrary units [see e.g. \onlinecite{bermejo93, mukherjee97, bermejo99}] and exclude the possibility of a ``complete'' (i.e. both in shape and absolute-scale intensity of the spectra) comparison between data and calculations.

The construction of appropriate models is however made rather complex by the quantum nature of the hydrogen liquids, which is highlighted by the relatively large values of the de Broglie thermal wavelength $\Lambda =h/ \sqrt{2 \pi M k_{\rm B} T} $, where $M$ is the molecular mass, $T$ is the temperature, and $k_{\rm B}$  and $h$ are the Boltzmann and Planck constants, respectively. Indeed the low mass and low temperature of these liquids makes $\Lambda$ comparable with the molecular dimensions, so that distinguishability of the particles is retained (and use of quantum statistics is unnecessary) but delocalization effects show up in the scattering properties, as evidenced even for the heavier deuterium molecule (see e.g.\ [\onlinecite{celli98}]). 

Therefore classical analytical or simulation methods are not appropriate to predict the center-of-mass (CM) single-molecule ({\it self}) and total ({\it self} plus {\it distinct}) dynamic structure factors of liquid H$_2$ and D$_2$. It is important to stress that the hydrogens are ``quantum'' liquids in the above special meaning, which is weaker than that of the He case, but is anyway more profound than that brought about by the non-commutative properties of position and momentum operators, leading to detailed balance spectral asymmetry and non-zero first frequency moment even in a non-interacting monatomic system. Furthermore, quantization of the internal degrees of freedom (e.g.\ rotations and vibrations) of molecular fluids is once again something else, which has to be duly considered also for ``classical'' liquids such as water or methane, and not only in the case of the hydrogens.    

Much work was devoted in past years to the implementation of scattering kernels for the hydrogen liquids \cite{MacFarlane94,Morishima94,Granada2003, Granada2004,Morishima2004} with different methods used to effectively evaluate both the self and distinct CM contributions to the DDCS. However, the above mentioned quantum effects were mostly neglected, or accounted for only approximately, or finally included in an effective way by adjustments to pioneering experimental data. \cite{egelstaff, carneiro, bermejo93, mukherjee97, bermejo99} 
 
In order to investigate their role in the neutron response of these liquids, we focus here on the most significant case of H$_2$ at liquid temperatures (i.e., para-H$_2$), which combines a stronger quantum behavior with a full predominance (in most kinematic conditions) of the self component, the latter being quite reliably accessed by quantum simulation methods (see Sect. III) and sensitively probed by neutrons in a unique way.

\section{The DDCS of H$_2$}

By assuming free rotations and neglecting rotation-vibration coupling, the nuclear neutron scattering by a homonuclear diatomic molecule can be schematically described by 

\begin{equation}
\label{d2sig} 
\frac{d^2\sigma}{d\Omega dE}=\sqrt{\frac{E_1}{E_0}} S_{\rm n}(Q,E) \nonumber
\end{equation}

\noindent with: \cite{guariniDDCS}

\begin{equation}
\begin{split}  
\label{Sneutrbiat} 
&S_{\rm n}(Q,E)= u(Q) S_{\rm CM,dist}(Q,E)+\\
&+ \sum_{J_0 J_1 v_1} F_{J_0 J_1 v_1}(Q) S_{\rm CM,self}(Q,E-E_{J_0 J_1}-E_{0 v_1}) 
%S_{\rm n}(Q,E)= u(Q) S_{\rm CM,dist}(Q,E)+\\
%+ \sum_{J_{0} J_{1} v_{1}} F(Q,J_{0}, J_{1}, v_{1}) S_{\rm CM,self}(Q,E-E_{J_{0} J_{1}}-E_{0 v_{1}}) 
\end{split}
\end{equation} 

\noindent where $u(Q)$ is a $Q$-dependent function containing only the coherent cross sections of the nuclei in the molecule, and the function $F$ takes different expressions according to the nuclear spin statistics and the ortho-para concentration, and contains both the coherent and incoherent nuclear cross sections (see [\onlinecite{guariniDDCS}] ). In Eq.\ (\ref{Sneutrbiat}) $S_{\rm CM,dist}(Q,E)$ and $S_{\rm CM,self}(Q,E)$ denote the distinct ($j \neq j^{\prime}$) and self ($j=j^{\prime}$) components of the total dynamic structure factor per molecule $S_{\rm CM}(Q,E)$:

\begin{equation}
\begin{split}  
\label{sqw} 
 S_{\rm CM}(Q,E)=\frac{1}{2\pi\hbar}\int_{-\infty}^{+\infty}dt\exp(-i \frac{E}{\hbar}
 t)\times\\
 \times\langle\frac{1}{N}\sum_{j,j^{\prime}=1}^N \exp(-i{\bf Q}\cdot{\bf 
 R}_j(0))\exp(i{\bf Q}\cdot{\bf R}_{j^{\prime}}(t))\rangle.
 \end{split}
\end{equation} 

\noindent where $N$ is the number of molecules and  ${\bf R}_j$ and ${\bf R}_{j^{\prime}}$ denote the CM position operators of the $j$-th and $j^{\prime}$-th molecule at time 0 and $t$, respectively. 

The second term of Eq.\ (\ref{Sneutrbiat}) represents the single-molecule dynamics $S_{\rm CM,self}(Q,E)$ convoluted with the line structure of the internal molecular motions, and results in a sum of spectral lines centered at the energies of rotational ($J_{0} \to J_{1}$) and vibrational ($v_{0} \to v_{1}$) transitions, where the subscripts 0 and 1 are used to label initial and final state, respectively. The ground vibrational state ($v_{0}=0$) is assumed here as the only one significantly populated in hydrogen at liquid temperatures.

As anticipated, the last term in Eq.\ (\ref{Sneutrbiat}) is particularly important in the case of liquid H$_2$ due to the huge incoherent-to-coherent cross section ratio of the hydrogen nucleus. Moreover, in the case of para-H$_2$ the combination of initial-state probabilities and spin correlations makes the DDCS spectra dominated, in most kinematic conditions, by the intense rotational line with $J_{0}=0$ and $J_{1}=1$, at $\sim$ 14.7 meV. \cite{Celli2000} Consequently, a good modeling of $S_{\rm CM,self}(Q,E)$ alone is generally sufficient to account for the scattering of thermal and hot neutrons from this liquid. In this respect, the results on liquid para-H$_2$ provide a fundamental test of the general quality of possible representations for the single-molecule dynamics.

The simplest algorithms for the evaluation of the single-molecule part of the DDCS    \cite{MacFarlane94,Granada2003,Granada2004, NJOY2012,schott, Utsuro} originate from direct or empirically modified use of, basically, two possible analytical models for $S_{\rm CM,self}(Q,E)$: either the ideal gas (IG) law (see e.g.\ [\onlinecite{guariniDDCS}]), giving rise to the well-known Young and Koppel model \cite{youngkoppel} for the self DDCS; or the Egelstaff and Schofield (ES) model, \cite{egelstaffschofield} modified to comply with detailed-balance asymmetry, \cite{egelstaffsoper} and with the first frequency moment sum rule that ensures translational spectra with a (non-zero) first moment equal to the recoil energy $E_{\rm r} = \hbar^{2} Q^{2} / 2 M$. By contrast, in Ref.\ [\onlinecite{Morishima94}] use is made of a semiempirical determination of the spectrum of the CM velocity autocorrelation function (VACF) of a hydrogen molecule, which, in the Gaussian approximation (GA), \cite{vineyard,RSS} is related to the intermediate scattering function $F_{\rm CM,self}(Q,t)$ and, through a time Fourier transform, to $S_{\rm CM,self}(Q,E)$. However, in Ref.\ [\onlinecite{Morishima94}] the GA was not applied in the rigorous quantum-mechanical form given, for example, in Ref.\ [\onlinecite{RSS}] and summarized in the next section.

Although different from each other, all the above approaches suffer anyway from the limitation of neglecting or approximating the quantum behavior of the liquid, in the sense we specified previously. Moreover, none of them provides direct agreement with experiment unless by modifying some property entering the model (like e.g., in the ES model, the mass of the molecule or the self diffusion coefficient) or by resorting to adjustable parameters in the spectrum of the VACF, as in the case of Ref.\ [\onlinecite{Morishima94}]). In some cases, TCS measurements have been fairly reproduced by considering the effect of intermolecular vibrations on the VACF spectrum and by hypothesizing the presence of huge molecular clusters diffusing in liquid hydrogen, leading to a $\sim$ 35-40 times augmented molecular mass \cite{egelstaff,egelstaffbook,Granada2003, Granada2004} to be used in the diffusive part (ES) of the single-molecule dynamics. However, such a high mass value, usually explained in terms of translational hindering and cluster formation, is not easily justified, and might reveal instead some inadequacy of the ES lines-hape.

These observations induced us to explore whether model parametrization and fit-based adjustments to experimental data could be fully avoided by means of a quantum determination of $S_{\rm CM,self}(Q,E)$. Moreover, this could help solving known physical inconsistencies, like the violation of the quantum second moment sum rule \cite{balucani} by the ES and IG spectral line-shapes. In addition to the compliance with the quantum properties of the liquid, other advantages of such a parameter-free and first-principle method are evident: i) a straightforward adaptation to different thermodynamic and kinematic conditions; ii) the possibility to avoid introducing not well-justified hypotheses and to use the true H$_2$ molecular properties; iii) good control of the physical consistency of the used line-shape for $S_{\rm CM,self}(Q,E)$ and of its compliance with the basic sum rules for a quantum fluid.

The next section illustrates the method and the successful achievement of the above expectations both for the spectral properties of $S_{\rm CM,self}(Q,E)$ and for the neutron TCS results at thermal and epithermal neutron energies.

\section{Quantum simulation-based self dynamics}

The rationale behind the attempt here described is given by the recent validation of quantum Centroid Molecular Dynamics (CMD) as an effective method to simulate the VACF of hydrogen and the detailed probing of the degree of accuracy (and range of applicability) of the Gaussian approximation in predicting the measured line-shape of liquid para-H$_2$. \cite{celli2011} Indeed, a non-Gaussian behavior was observed in an intermediate $Q$ range. However, the differences between neutron data and calculations, combining CMD simulations with the GA (denoted in the following as CMD+GA), are small enough to suggest that they may be irrelevant for cross-section calculation purposes, especially at the level of doubly integrated quantities like the TCS.

As described in Ref. [\onlinecite{celli2011}], the Path Integral CMD method was applied to a system of 256 molecules interacting via the Silvera-Goldman potential \cite{silvera}. The Trotter number, i.e. the number of beads on the classical ring polymers replacing the quantum mechanical particles, was 64. In contrast to the usual implementation, the calculation of the quantum mechanical forces, which are required at each time step of the otherwise classical simulation, was performed by the path integral Monte Carlo method, rather than MD, thus avoiding sampling problems associated with the stiff ``intramolecular” modes of the polymers and allowing for a much larger time step. The simulation was extended up to 1 ns in the isokinetic ensemble, ensuring thermal stability and statistical reliability. The velocity correlation was calculated up to a maximum time lag of 1.5 ps. A shorter test run with 500 particles confirmed that the shape of the VACF was not noticeably influenced by finite-size effects. 

The dynamical information conveyed by the VACF is a keypoint in the development of models for the self part of the DDCS of viscous dense fluids. In particular, the output of a PI CMD simulation is the canonical (or Kubo-transformed \cite{kubo}) VACF: 
 
\begin{equation}  
\label{kubo_VACF} 
u_{c}(t) = {k_{\rm B} T} \int_{0}^{\frac{1}{k_{\rm B} T}}{d\lambda ~ \Bigg<e^{\lambda H}{\bf v}_ {\rm CM}(0) ~\cdot ~e^{-\lambda H}{\bf v}_ {\rm CM}(t)\Bigg>} 
\end{equation} 

\noindent where $H$ is the Hamiltonian operator of the system. The self intermediate 
scattering function $F_ {\rm CM,self}(Q,t)$ in the Gaussian approximation can then be written as \cite{RSS}: 
 
\begin{equation}  
\begin{split}
\label{Fself} 
F_{\rm CM,self}(Q,t) =\exp\Bigg[ - \frac{E_{\rm r}}{\hbar} \int_{0}^{+\infty}{d\omega} ~~\frac{f(\omega)}{\omega} ~A(\omega)\Bigg],
\end{split}
\end{equation} 
 
\noindent with 

\begin{equation}  
\begin{split}
\label{gamma} 
\nonumber
A(\omega)= [1-\cos(\omega t)] \coth \Bigg( \frac{\hbar \omega}{2 k_{\rm B} T} \Bigg) - i \sin (\omega t)   
\end{split}
\end{equation} 

\noindent and

\begin{equation}  
\label{fom} 
\nonumber
f(\omega)= \frac{M}{3 \pi k_{\rm B} T}  \int_{-\infty}^{+\infty}{dt ~e^{-i \omega t} ~u_{c}(t)}. 
\end{equation} 

\begin{figure} 
\includegraphics[angle=0,width=1\textwidth]{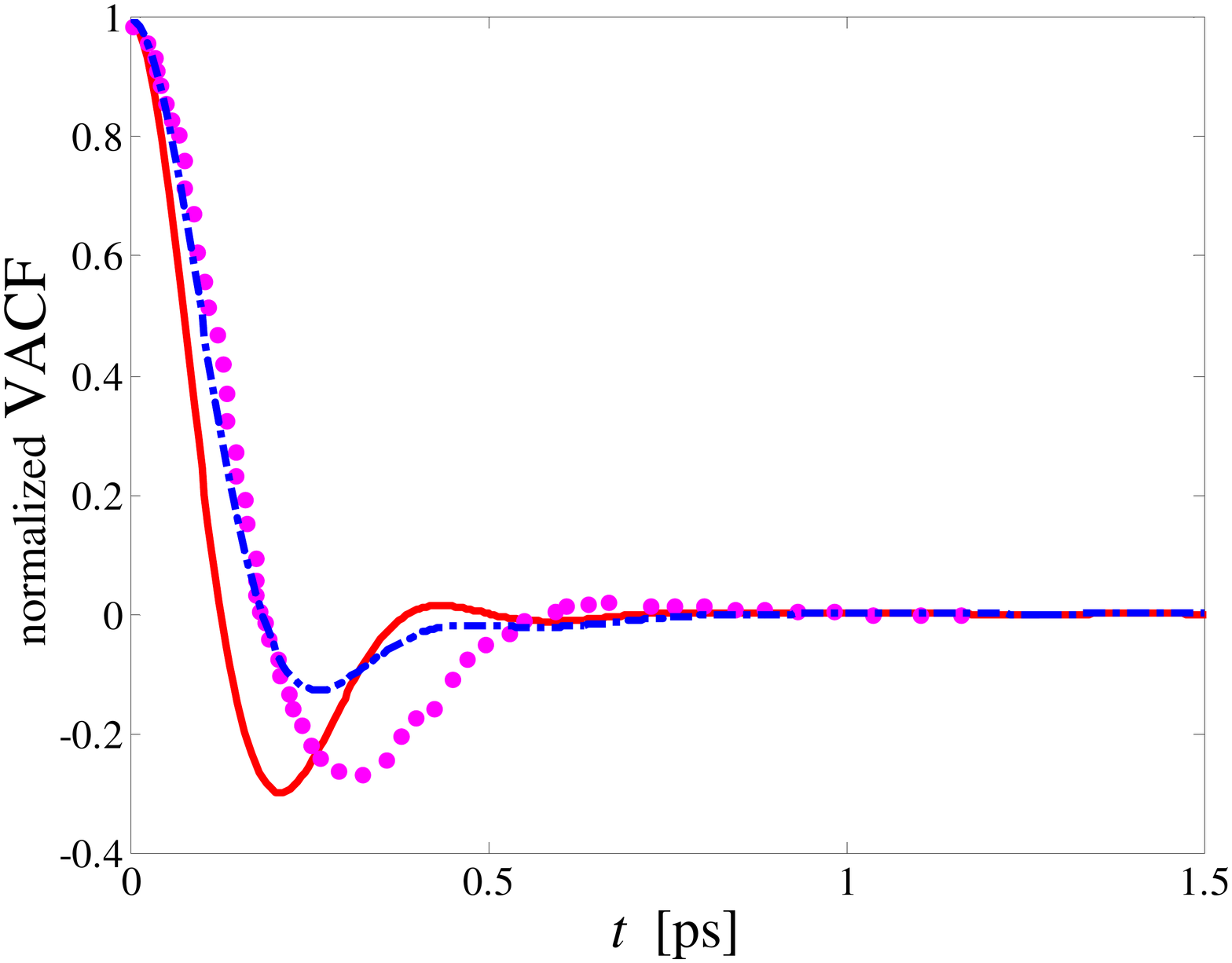} 
\caption{(Color online) Velocity autocorrelation function of hydrogen deduced in [\onlinecite{Morishima94}] (pink dots). The quite different results of the quantum CMD simulations \cite{celli2011} are also reported. For comparison, both the canonical VACF $u_{c}(t)$ (dash-dotted blue curve) and the real part of $ \langle{\bf v}_ {\rm CM}(0) \cdot {\bf v}_ {\rm CM}(t)\rangle$ (red solid curve) are shown.} 
\label{VACF} 
\end{figure}

\noindent Figure \ref{VACF} reports our simulation results for the VACF of para-H$_2$ at 15.7 K: both the canonical VACF $u_{c}(t)$ and the real part of $ \langle{\bf v}_ {\rm CM}(0) \cdot {\bf v}_ {\rm CM}(t)\rangle$ are displayed for comparison with the VACF values digitalized from Fig. 3 of Ref. [\onlinecite{Morishima94}], also shown in the figure. The differences between the present CMD results and the past semi-empirical determination of Ref. [\onlinecite{Morishima94}] at 14.7 K seem too large to be ascribed to a mere temperature effect.  

\begin{figure} 
\includegraphics[angle=0,width=1\textwidth]{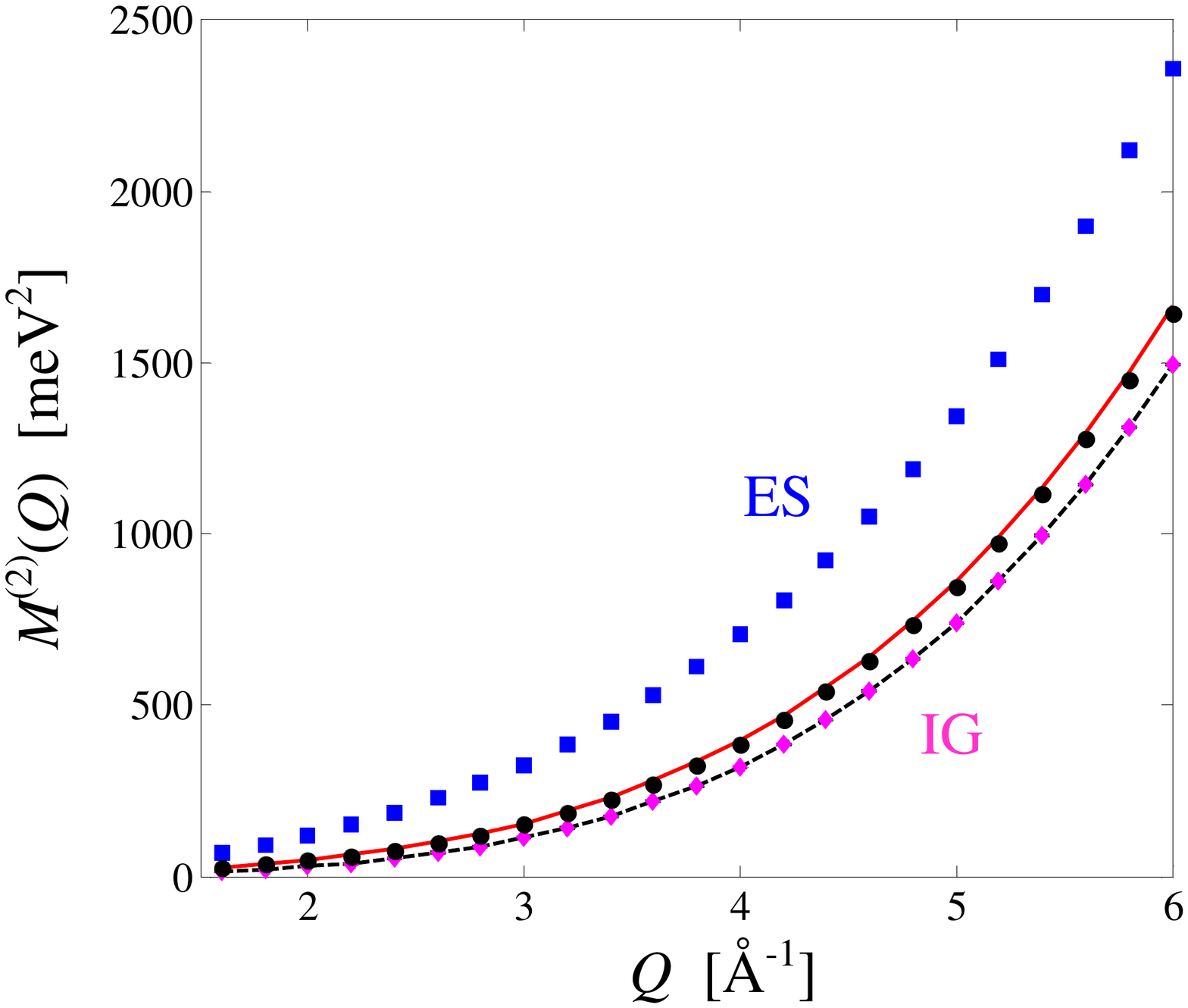} 
\caption{(Color online) Second frequency moment of para-H$_2$ at 15.7 K calculated from Eq. (\ref{q_sec_mom}) either using the experimental $ \langle E_{\rm K} \rangle$ estimate of Ref. [\onlinecite{celli2002}] (red solid curve), or the 
classical mean kinetic energy (3/2) $k_{\rm B} T$ (black dashed curve). Symbols correspond to the values obtained instead by energy integration of the IG (pink diamonds), the ES (blue squares), and the CMD+GA (black dots) spectra.} 
\label{mom2conGA} 
\end{figure}

\noindent The self dynamic structure factor, as a function of $\omega=E/\hbar$, is then obtained as the time Fourier transform of Eq. (\ref{Fself}) at each desired $Q$. By integration of the spectra over very wide energy ranges, it is possible to verify the consistency with the second moment sum rule which, for a quantum system, reads:\cite{balucani}

\begin{equation}  
\begin{split}
\label{q_sec_mom} 
M^{(2)}(Q)= \int{dE ~E^2  ~S_{\rm CM,self}(Q,E)}= \\
=\frac{2 \hbar^{2} Q^2}{3M} \langle E_{\rm K}\rangle +E_{\rm r}^2,  
\end{split}
\end{equation} 

\noindent where $ \langle E_{\rm K} \rangle$ is the mean kinetic energy of the particle, which in the present case differs significantly from the classical value (3/2) $k_{\rm B} T$. Experimental and Path Integral Monte Carlo simulation values of $\langle E_{\rm K} \rangle$ for para-hydrogen at various liquid temperatures have been provided by Celli et al. \cite{celli2002} and Colognesi et al.. \cite{colognesi2004} Fig. \ref{mom2conGA} shows the $Q$-dependence of the theoretical prescription of Eq. (\ref{q_sec_mom}) (lines) in comparison with the IG, ES and CMD+GA results for the second frequency spectral moment (symbols). In particular, the continuous red curve corresponds to the calculation of Eq. (\ref{q_sec_mom}) using the experimental value of the mean kinetic energy of para-H$_2$ at 15.7 K, \cite{celli2002} while the dashed black curve is derived by assuming a classical mean kinetic energy of (3/2) $k_{\rm B} T$. As expected, the latter calculation agrees very well with the values obtained by appropriate energy integration of the IG spectra. Similarly, the CMD+GA values are in very good agreement with the quantum behavior. Conversely, the ES model misses both the classical and quantum prescriptions of $M^{(2)}(Q)$.

\section{Comparison with experimental data in the thermal and epithermal range}

\begin{figure}
\resizebox{0.48\textwidth}{!}{%
\includegraphics[viewport=2.5cm 1cm 25.5cm 21cm]{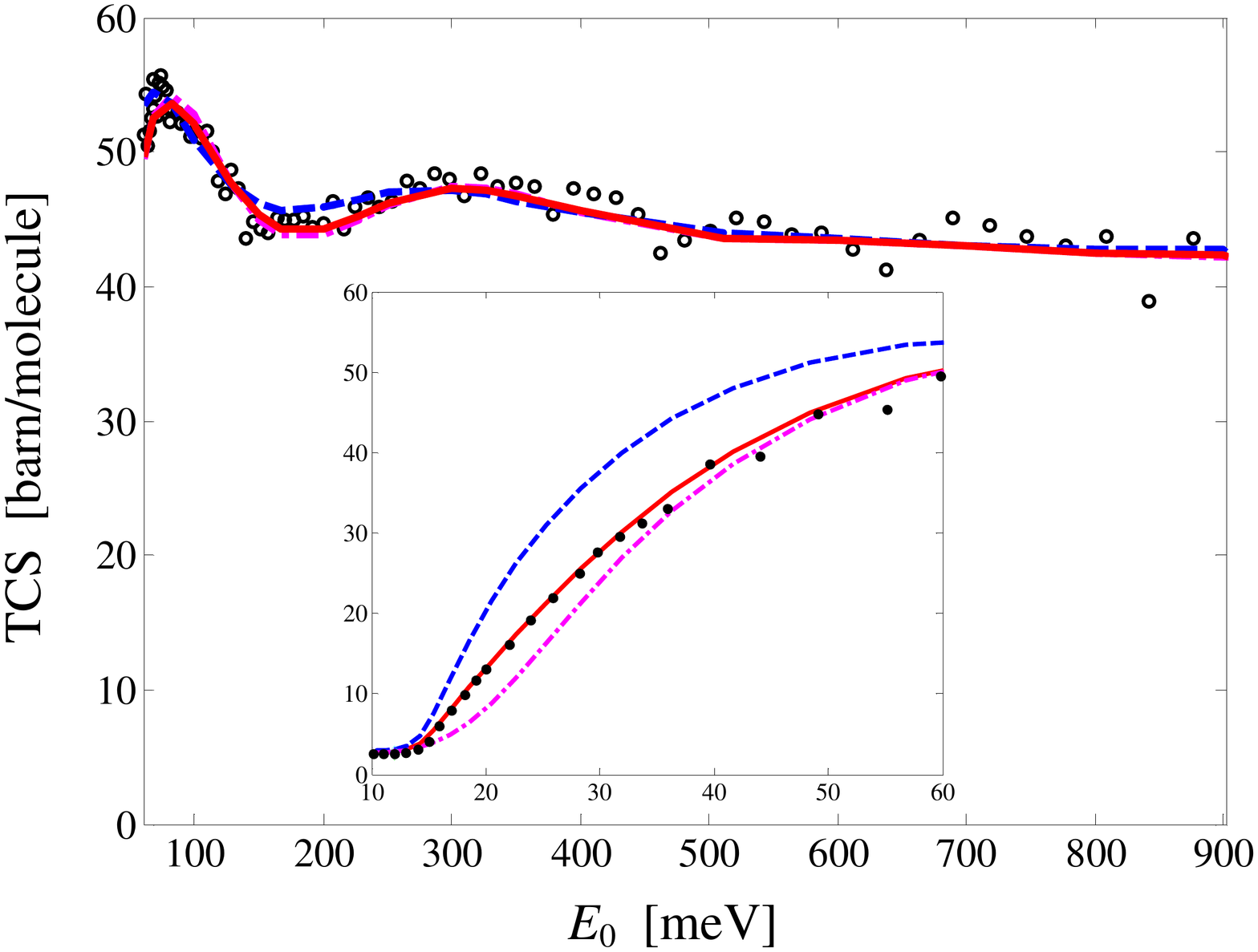}
}
\caption{(Color online) Total cross section per molecule of para-H$_2$ at 15.7 K and 1 atm, at the thermal energies of Seiffert measurements \cite{seiffert} (black dots in the inset) and at the higher $E_0$ values probed in the spallation-source experiment of Celli et al.\ \cite{celli1999} (black empty circles in the main frame). Experimental data are compared with the IG (pink dash-dotted line), ES (blue dashed line) and CMD+GA (red solid line) calculations.} 
\label{TCSH2} 
\end{figure}

The previous results show that the CMD+GA $S_{\rm CM,self}(Q,E)$ has the correct spectral properties and can sensibly be used to perform the self DDCS estimates based on the computation of the second term of Eq.\ (\ref{Sneutrbiat}), which we implemented according to the details given in Ref. [\onlinecite{guariniDDCS}]. By numerical integration, over energy and solid angle, of the DDCS spectra we finally derived the TCS values reported in Fig. \ref{TCSH2} which, as mentioned, regard an incident energy range where the distinct contributions to Eq.\ (\ref{Sneutrbiat}) are fully negligible to a good approximation. Calculations were carried out also using the IG and ES line-shapes for comparison with the quantum simulation-based outputs. From the inset of Fig.\ \ref{TCSH2} the superiority of the CMD+GA results in reproducing the experimental TCS of para-H$_2$ at thermal incident energies is evident. At higher energies, as expected, the system tends to ideal gas behavior: all calculated curves thus become indistinguishable and in very good agreement with the experimental data.  

\begin{figure*}
\resizebox{0.98\textwidth}{!}{%
\includegraphics[viewport=1cm 5cm 29cm 14cm]{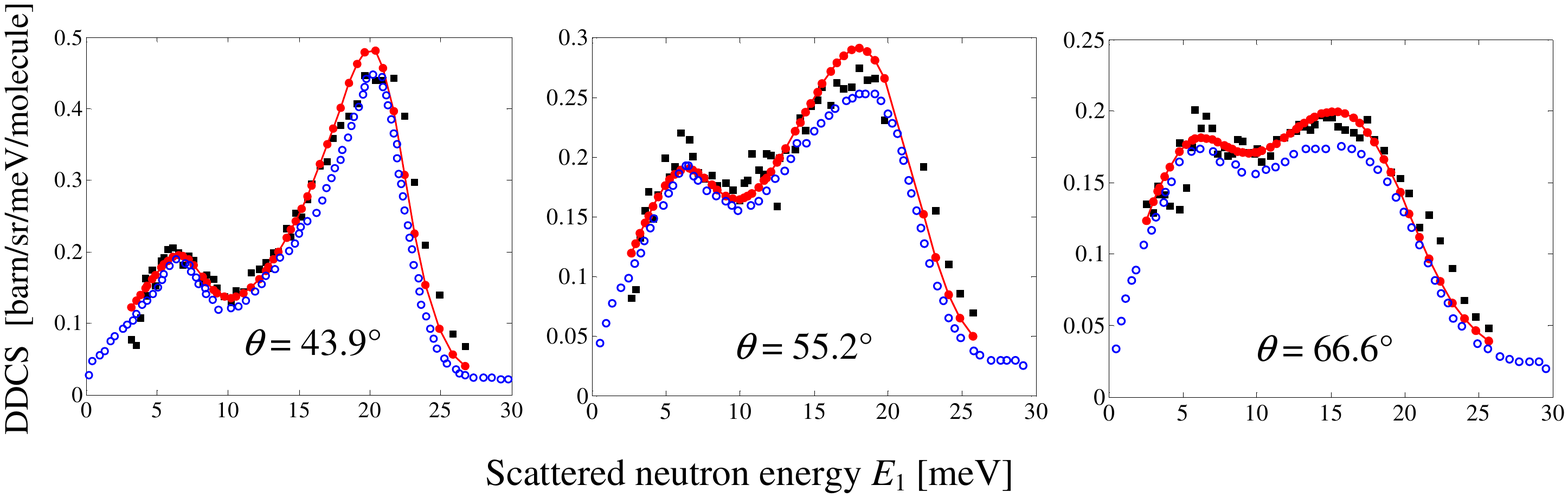}
}
\caption{(Color online) Experimental DDCS of H$_2$ at 19.8 K with a 41\% ortho concentration, (black squares) \cite{schott} and CMD+GA results (red dots with thin line). The blue empty circles are the DDCS values calculated by Morishima and Nishikawa. \cite{Morishima2004} Measurements refer to the thermal range ($E_0$ = 21.8 meV) and rather wide angles indicated in each frame.} 
\label{DDCS_schott} 
\end{figure*}

At the more detailed level of non-integrated quantities, the effectiveness of the CMD+GA method can be partially tested against one of the few inelastic scattering measurements \cite{schott} providing data in absolute units. In this case, since the hydrogen sample had a 41\% ortho concentration, the CMD+GA computations were performed in the same conditions. The CMD+GA calculated spectra satisfactorily describe the experimental data of Fig.\ \ref{DDCS_schott}. This shows that use of a quantum representation of $S_{\rm CM,self}(Q,E)$ is the only real requirement to obtain a direct and reasonable agreement with experiment, comparable with the one obtained with the calculations of Ref.\ [\onlinecite{Morishima2004}], and incomparably better than that of Schott himself (using the IG model) and of Utsuro \cite{Utsuro} (both not shown in the figure). 

The method of combining VACF quantum simulations with the GA represents therefore an extremely valid alternative to most experiments or to the use of ``quantum-insensitive'' analytical models for $S_{\rm CM, self}(Q,E)$ in the DDCS formula of H$_2$. A great advantage of this method, recently applied with simpler classical simulations to the case of water, \cite{abe} is that the VACF only depends on the thermodynamic state, and only one simulation run is required to enable, in a given state, calculations of $S_{\rm CM, self}(Q,E)$ wherever wished in the kinematic ($Q,E$) plane. This actually means that, at least for H$_2$, we are presently able to calculate the DDCS in most kinematic conditions, with good accuracy, and high control on the physical consistency of the results. 

\section{Comparison with experimental data in the subthermal range}

Differently from the case of more energetic incident neutrons, at cold neutron energies ($1 < E_0 < 10$ meV) the interplay of initial-state probabilities and spin correlations in para-hydrogen actually kills out the incoherent signal, leaving the small coherent cross section at the rudder of both collective and single-molecule contributions. Indeed, if only quasielastic scattering is allowed (because cold neutrons are unable to induce transitions in para-H$_2$) and only the $J_0=0$ level is thermally populated (because of the low temperature of this liquid) it can be shown that the second term of Eq.\ (\ref{Sneutrbiat}) reduces to $u(Q) S_{\rm CM, self}(Q,E)$, so that $S_{\rm n}(Q,E)= u(Q) S_{\rm CM}(Q,E)$. As a result, at neutron energies below the threshold of the first rotational transition for this system, the distinct dynamics warrants a primary role also in the response from a nominally "incoherent" liquid as para-H$_2$. Actually, there are evidences that distinct intermolecular contributions influence the neutron signal from para-H$_2$ and that self calculations are insufficient in certain conditions.\cite{Morishima2004}

A good knowledge of the collective (i.e.\ coherent) dynamics of this liquid is therefore an essential ingredient to achieve the accuracy demanded nowadays on subthermal neutron cross sections. Unfortunately, present quantum simulations methods are not yet able to provide direct and reliable estimates of the self and total (self plus distinct) dynamic structure factors. \cite{Jang,Voth} At the same time, no analytical model exists for the total $S_{\rm CM}(Q,E)$ of a quantum liquid, and more generally of any liquid, except that in hydrodynamic ($Q \to 0$) conditions.
Therefore, we could only try to investigate, in an approximate way, the effect of adding a collective term to the single-molecule results. To do this, we adopted the   Sk\"{o}ld approximation, \cite{Skold} that models the total $S_{\rm CM}(Q,E)$ through a modification of its self part, namely:

\begin{equation}  
\label{skold} 
S_{\rm CM}(Q,E) \approx S_{\rm CM}(Q) S_{\rm CM,self}\Bigg(\frac{Q}{\sqrt{S_{\rm CM}(Q)}},E\Bigg),
\end{equation} 

\noindent which was cleverly conceived to fulfill the second moment sum rule in classical systems, and gave satisfactory results in several cases, starting from that of liquid argon. \cite{Skold2} In Eq.\ (\ref{skold}) $S_{\rm CM}(Q)$ is the CM static structure factor for which we took the experimental values obtained by neutron diffraction measurements. \cite{celli2005}

\begin{figure} 
\includegraphics[angle=0,width=1\textwidth]{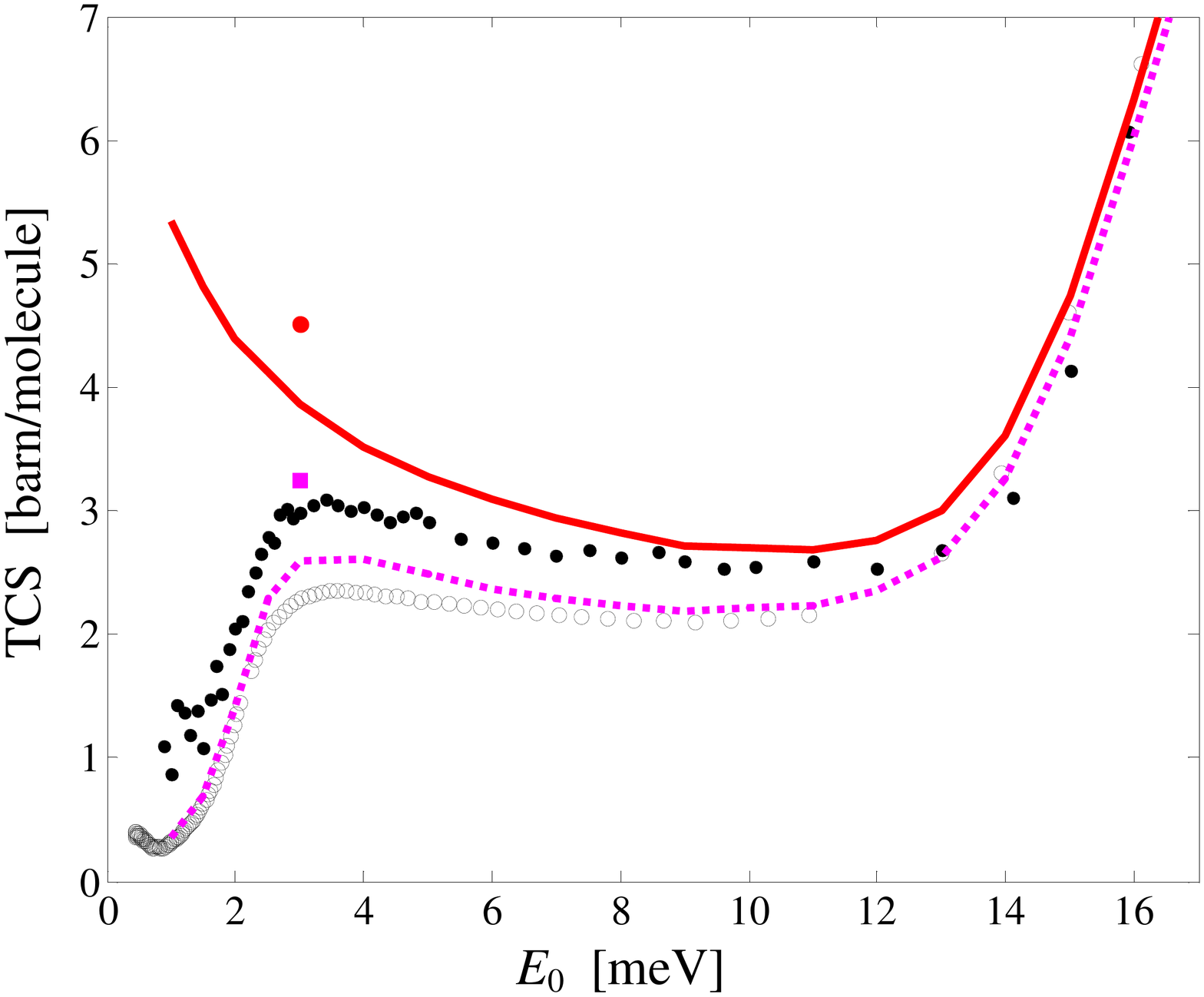} 
\caption{(Color online) Total scattering cross section of para-H$_2$ at cold neutron energies. The large deviation of the CMD+GA self contribution (red solid curve) from the absorption-corrected experimental data of Seiffert \cite{seiffert} (black dots) and of Grammer {\it et al.} \cite{grammer2015} (black empty circles) is due to the missing negative distinct contribution. The pink dotted curve displays the results obtained by adding a distinct part in the Sk\"{o}ld approximation. The dots a 3 meV show the effect, on the self (red dot) and on the self+Sk\"{o}ld (pink square) TCS, of the presence of 0.5 $\%$ ortho-hydrogen.} 
\label{H2_lowE} 
\end{figure}

As evident in Fig. \ref{H2_lowE}, the CMD+GA calculation of the self part progressively departs from the TCS measurements presently available \cite{seiffert,grammer2015}, as soon as the incident neutron energy is decreased below 14 meV. The Sk\"{o}ld schematization of the total dynamics, calculated by inserting in Eq. (\ref{skold}) the properly $Q$-scaled CMD+GA self dynamic structure factor, is seen instead to be rather effective, providing a suitable correction of the self-only results. In particular, despite its simplicity, the Sk\"{o}ld model, if used in combination with a quantum $S_{\rm CM,self}(Q,E)$, is able to quite satisfactorily describe the very recent neutron transmission data of Grammer and co-workers at 15.7 K, with only a small overestimate between 2 and 8 meV. Our quantum-based calculations thus apparently confirm the lower cross section values of liquid para-H$_2$ with respect to the earlier cold neutron determinations by Seiffert. A possible contamination, by a small fraction (0.5 $\%$) of ortho molecules, of Seiffert sample has been plausibly hypothesized in Ref. [\onlinecite{grammer2015}]. This actually seems to be the case, as shown in Fig. \ref{H2_lowE}, where we also report, for an example incident energy of 3 meV, the changes induced, in both the CMD+GA and Sk\"{o}ld TCS, by the presence of  0.5 $\%$ ortho-H$_2$.  Indeed, temperature effects might also be thought of to find an explanation to the quite large discrepancies between the two TCS data-sets (15.7 K against 14 K). However, it is well known that structural changes are extremely limited with varying temperature in high density liquids, \cite{guariniD2} and could unlikely lead to such significant differences in the TCS.

%Indeed, looking at Sk\"{o}ld formula, the $S(Q)$ variation with $T$ is the steering quantity that mainly determines, after integration, the changes in the TCS, since the spectral shape of $S_{\rm CM,self}(Q,E)$ does not vary significantly  .   

The origin of the residual differences between Grammer {\it et al.} data and our calculations in the range 2$<E_0<$8 meV, cannot be easily found out. They might either depend on inaccuracies of the data, or of the $S(Q)$ determination at the $Q$ values involved at certain incident energies, or finally on both. It is also plausible that the featureless line-shapes produced by the Sk\"{o}ld recipe do not provide an adequate description of the total dynamic structure factor which, besides the single-molecule dynamics, accounts for collective excitations as well, with well-known inelastic features. At low and intermediate wave vectors  (e.g. up to that of the main peak of $S(Q)$), substantial spectral differences from the Sk\"{o}ld model are therefore expected, and might reveal themselves in a constant-$\theta$ energy integration between (ideally) $-\infty$ and $E_0$, like the one performed to compute the TCS. Finally, another possible explanation of the small deviations could be a slight inadequacy of the CMD+GA method, and in particular a signature of the mentioned non-Gaussian behavior of para-H$_2$ in certain $Q$ ranges. In this respect, some observations are worth. Although deviations from Gaussian behavior were found to be rather small at the level of frequency spectra,\cite{celli2011} the effect of even small discrepancies has a different impact on the TCS results depending on $E_0$, and is more important if the $Q$ values where the GA fails belong, in majority, to the explored kinematic region. Conversely, at very low $E_0$ (e.g. 1-2 meV) the probed $Q$ region predominantly lies between values where the GA holds, explaining the renewed agreement, as energy is decreased below 2 meV. Finally, at higher $E_0$ (e.g. between 8 and 14 meV) many, rather low and rather high, $Q$ values (where the GA holds) contribute to the TCS, so the role of the ``GA-deviating'' $Q$ values has a much lesser impact on the integrated result, and agreement with experiment is found again. In our opinion, this might be a sensible explanation of the initially growing and then decreasing discrepancies with increasing $E_0$ in the subthermal range.  Anyway, it is noteful that the present CMD+GA approach allows to reach, using in the calculations the true molecular properties of para-H$_2$, such a satisfactory description of the new measured data.

Clearly, these are plausible speculations that can only be verified by further research. Two different routes can be envisaged in order to understand, if possible, the role of the two main approximations here adopted (i.e., GA and Sk\"{o}ld) to calculate the subthermal TCS. One is the introduction in our DDCS algorithm of the first non-Gaussian correction to the self dynamics. \cite{nonGauss} Though feasible, such an attempt is certainly ``expensive'' from the computational point of view, but it is also extremely attractive in a scientific sense, owing to the great opportunity to test possible (near to ``macroscopic'') failures of the GA, in case they show up even at the level of neutron TCS calculations.  In addition, the dependence on $E_0$ of the supposed GA inadequacy could be duly checked. The other route passes through experiments aimed at determining the dynamic structure factor and comparing it with the Sk\"{o}ld prediction. Indeed, accurate measurements of the DDCS of liquid para-H$_2$, used in combination with the available simulations of the self part, are needed to be able to sensitively probe the different contributions to the total signal in the sub-thermal and cold incident energy range below $\sim$ 14 meV, with a great effort in producing well-normalized and duly corrected scattering data. 

\section{Conclusions}

We showed that it is now possible to obtain accurate evaluations of the thermal neutron scattering law of cryogenic liquids as important as para-H$_2$. We developed, verified and implemented an efficient simulation-based method able to accurately account for the quantum behavior of the fluid, and which has the doubtless merit of limiting considerably the need of consuming experiments, at least as regards the thermal and epithermal range. The whole kinematic plane can indeed be covered by this technique, avoiding demanding experiments and with full flexibility. First-principle methods, use of the known molecular properties of hydrogen, and neither search of forced agreement with experiment nor introduction of effective mass values or other parametrized routes, have been successfully experimented. An improved evaluation of the non-Gaussian behavior of para-H$_2$ is suggested by the present TCS results, which point also at the importance of accurate measurements of the DDCS at some subthermal incident neutron energies where the role of the distinct contribution is, as well, worth investigating.
The present estimates represent, to our knowledge, the only ones providing a parameter-free absolute-scale agreement with total cross section measurements, ensuring, at the same time, compliance with the spectral quantum sum rules. In this sense, the results reported in this work also constitute a rather convincing validation, at present, of the CMD simulation technique for the prediction of the VACF of quantum liquids. This also highlights the importance of quantum simulation research, both for fundamental motivations, and for its fruitable use in increasingly demanding fields of application, like the management and new-concept development of neutron sources.

% If in two-column mode, this environment will change to single-column format so that long equations can be displayed.
% Use only when necessary.
%\begin{widetext}
%$$\mbox{put long equation here}$$
%\end{widetext}

% Figures should be put into the text as floats. 
% Use the graphics or graphicx packages (distributed with LaTeX2e).
% See the LaTeX Graphics Companion by Michel Goosens, Sebastian Rahtz, and Frank Mittelbach for examples.
%
% Here is an example of the general form of a figure:
% Fill in the caption in the braces of the \caption{} command. 
% Put the label that you will use with \ref{} command in the braces of the \label{} command.
%
% \begin{figure}
% \includegraphics{}%
% \caption{\label{}}%
% \end{figure}

% Tables may be be put in the text as floats.
% Here is an example of the general form of a table:
% Fill in the caption in the braces of the \caption{} command. Put the label
% that you will use with \ref{} command in the braces of the \label{} command.
% Insert the column specifiers (l, r, c, d, etc.) in the empty braces of the
% \begin{tabular}{} command.
%
% \begin{table}
% \caption{\label{} }
% \begin{tabular}{}
% \end{tabular}
% \end{table}

% If you have acknowledgments, this puts in the proper section head.
\begin{acknowledgments}
E. G. gratefully acknowledges the precious support of the whole ILL staff during her 2014 stay in Grenoble. The authors acknowledge the courtesy and extremely kind collaboration of K. B. Grammer and co-workers for having transmitted their results even prior to publication.
\end{acknowledgments}

% Create the reference section using BibTeX:

%\bibliography{your-bib-file}

 %``” 

\end{document}